\documentclass[a4paper, oneside, twocolumn, notitlepage, 10pt]{extarticle_ecoc}

\usepackage{ecoc}
\renewcommand\footnotemark{\textsuperscript{\S}}
\begin{document}
\selectlanguage{english}    


\title{Record Photon Information Efficiency with Optical Clock Transmission and Recovery of 12.5 bits/photon over an Optical Channel with 77 dB Loss}%

\author{
    Ren\'{e}-Jean Essiambre\textsuperscript{(1)}, Cheng Guo\textsuperscript{(1,2)}, Sai Kanth Dacha\textsuperscript{(1)},\footnote{\textsuperscript{\S}The contributions of S.K. Dacha, A. Blanco-Redondo, M. Weiner and N. Menkart were performed while working at Nokia Bell Labs.}, Alexei Ashikhmin\textsuperscript{(1)},\\ Andrea Blanco-Redondo\textsuperscript{(1),\S}, Frank R. Kschischang\textsuperscript{(3)}, Konrad Banaszek\textsuperscript{(4)}, Matthew Weiner\textsuperscript{(1),\S},\\ Rose Kopf\textsuperscript{(1)}, Ian Crawley\textsuperscript{(1)}, Mohamad H. Idjadi\textsuperscript{(1)}, Ayed A. Sayem\textsuperscript{(1)}, Jie Zhao\textsuperscript{(1)},\\ James D. Sandoz\textsuperscript{(5)}, Nicolas Fontaine\textsuperscript{(1)}, Nicole Menkart\textsuperscript{(1),\S}, Roland Ryf\textsuperscript{(1)}, John Cloonan\textsuperscript{(5)},\\ Michael Vasilyev\textsuperscript{(2)}, Thomas E. Murphy\textsuperscript{(6)}, Ellsworth C. Burrows\textsuperscript{(1)}}

\maketitle                  
\vspace*{-2cm}
\begin{strip}
 \begin{author_descr}

\textsuperscript{(1)} Nokia Bell Labs, 600 Mountain Ave, New Providence, NJ 07974, USA,

\textcolor{blue}{\uline{rene.essiambre@nokia-bell-labs.com}} 

\textsuperscript{(2)} Department of Electrical Engineering, University of Texas at Arlington, Arlington, TX 76019, USA

    \textsuperscript{(3)} Dept. of Electrical and Computer Engineering, University of Toronto, Toronto, ON M5S 3G4, Canada
    
    \textsuperscript{(4)} Centre for Quantum Optical Technologies, University of Warsaw, 02-097 Warszawa, Poland
    
   \textsuperscript{(5)} Nokia, 600 Mountain Ave, New Providence, NJ 07974, USA

   \textsuperscript{(6)} Institute for Research in Electron. and Appl. Phys., Univ. of Maryland, College Park, MD 20742, USA

 \end{author_descr}
\end{strip}

\setstretch{1.1}
\renewcommand\footnotemark{}
\renewcommand\footnoterule{}


\vspace*{-2cm}
\begin{strip}
  \begin{ecoc_abstract}
    We experimentally demonstrate optical detection at 12.5~bits per incident photon, 9.4~dB higher than the theoretical limit of conventional coherent detection. A single laser transmits both data and optical clock, undergoes 77~dB of attenuation before quantum detection followed by optical clock and data recovery. 
 \textcopyright2023 The Author(s)
  \end{ecoc_abstract}
\end{strip}

\section{Introduction}
The Photon Information Efficiency (PIE)\cite{gordon1962quantum} determines the maximum channel path loss and minimum signal power supported by optical communication systems. The PIE measures the information per unit of optical signal power incident on a receiver and is generally expressed in Bits per Incident Photon (BIP). An important application of high-PIE communication is space exploration\cite{sun2013free,williams2007rf}. A significant benefit of free-space optical over microwave communication is a $\sim$76-dB lower path diffraction loss\cite{caplan2007laser} due to nearly four orders of magnitude higher frequencies of optical waves ($\sim$200~THz) relative to microwaves ($\sim$30~GHz). Optical communication also offers considerably larger frequency bandwidth most of it unregulated\cite{toyoshima2021recent}.

A space optical communication project funded by the National Aeronautics and Space Administration (NASA) has demonstrated 22 (38) Mb/s with PIE of 0.28 (0.67) BIP over an Earth-Moon distance in~2015\cite{grein2015optical}. An optical communication technology demonstration reaching much farther is planned for the Psyche mission. It aims to optically communicate at 114 kbits/s with a PIE of 1 BIP at twice the maximum Earth-Mars distance\cite{biswas2018deep}. The launch is scheduled for October 5, 2023. 
\begin{figure*}[t]
   \centering
    \includegraphics[width=1\linewidth]{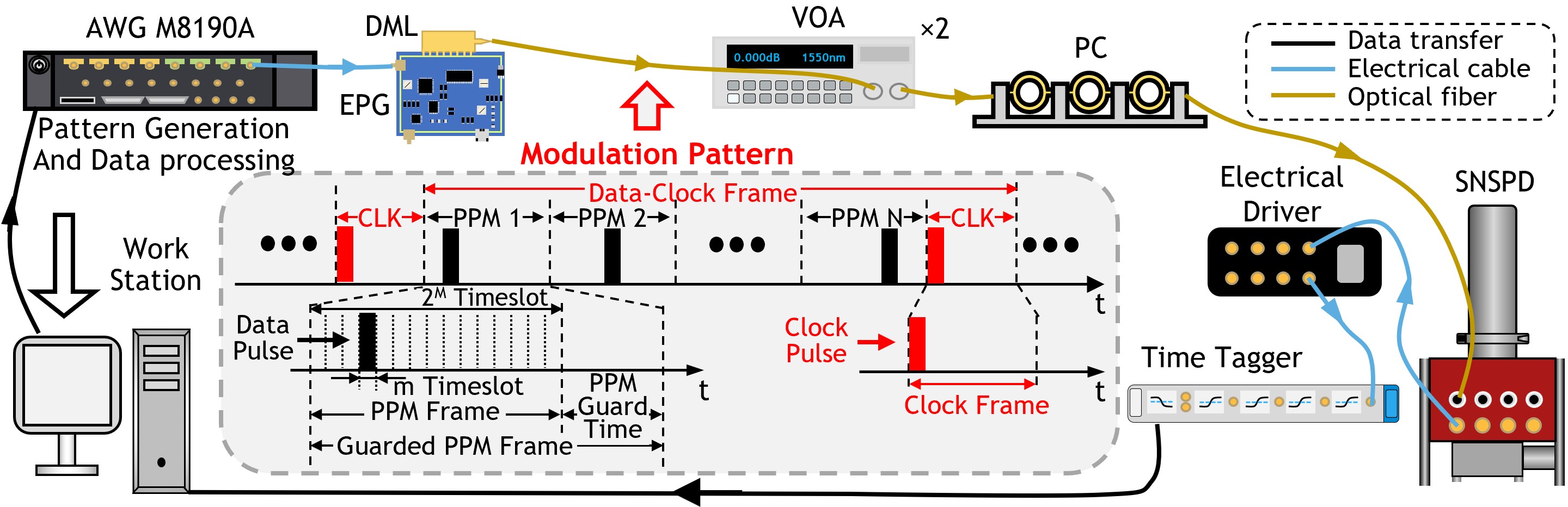}
    \caption{Experimental setup. AWG: Arbitrary Waveform Generator; CLK: Clock; DML: Direct Modulation Laser; EPG: Electrical Pulse Generator; PC: Polarization Controller; PPM: Pulse Position Modulation; SNSPD: Superconducting Nanowire Single-Photon Detector; VOA: Variable Optical Attenuator}
    \label{fig:figure1}
\end{figure*}
On Earth, the PIE achieved using an optical pre-amplifier and coherent receiver has been demonstrated to be as high as 0.48 BIP \cite{geisler2013demonstration} with phase-insensitive amplifier and 1.25 BIP with phase-sensitive amplifier\cite{kakarla2021power}. Without optical pre-amplifiers, a PIE of 0.67 BIP with a coherent receiver has been achieved\cite{stevens2008optical}. It was predicted as early as 1962 that photon counters could offer higher PIE than coherent detection\cite{gordon1962quantum}. Experimental demonstrations were performed using Geiger-mode avalanche photodiodes\cite{robinson20051} and Superconducting Nanowire Single-Photon Detectors (SNSPDs)\cite{craiciu2023high} and Pulse-Position Modulation (PPM)\cite{caplan2007laser}. With SNSPDs, the highest PIE demonstrated is 13.5 BIP at a rate of 6.4 kbits/s\cite{farr201313}. However, the same electrical clock was shared between transmitter and receiver, requiring them to be in close proximity and sidestepping the important issue of independent clocks synchronization, particularly challenging at high PIE and low received powers.

In this paper, we experimentally demonstrate an achievable PIE of 12.5 bits per incident photon at a data rate of 13.94 kbits/s with different free-running clocks at the transmitter and receiver over a channel with $77$~dB optical attenuation. To the best of our knowledge, this is the lowest energy-per-bit detection system operating with a fully independent transmitter and receiver. 

\section{Experimental setup}
Figure~\ref{fig:figure1} shows the experimental setup including a depiction of the modulation scheme. The {\it PPM frames} (or PPM symbols) are created by generating a single optical pulse in the position $m$ out of the $2^M$ PPM timeslots that are labeled from 1 to $2^M$ in increasing order of time. The location $m$ is given by the decimal representation of a segment of length $M$~bits of a pseudo-random binary sequence (PRBS). We run separate experiments with three PPM frame lengths of $M = 17$, $18$ or $19$. All frames and guard times are composed of an integer number of timeslots of 400~ps duration. 

The {\it PPM guard time} is set to 100~ns, i.e., 250 empty timeslots, appended after each PPM frame to form a {\it guarded PPM frame} that prevents detection blocking between PPM frames due to the 60-ns dead time of the SNSPDs\cite{moision2011blocking}. After $N$ guarded PPM frames, a 100~ns {\it clock frame} is inserted consisting of a single optical pulse in the first timeslot, followed by 249 empty timeslots. The ratio $N$ between the number of guarded PPM to clock frames is referred to as the data-clock ratio. For $N=1$, the clock frames doubles the signal power, causing a 3-dB penalty in PIE, while also reducing the PPM frame rate slightly (by 0.2\% or less in our experiments). For $N=10$, the 10\% power increase corresponds to 0.4-dB PIE penalty. The number of PPM frames recorded are 760k, 380k, and 190k, with 1600, 800, and 400 being distinct due to the memory limit of the arbitrary waveform generator (AWG), for PPM frame lengths of $2^{17}$, $2^{18}$ and $2^{19}$, respectively.\\
The Electrical Pulse Generator (EPG) modulates a 1550-nm Directly Modulated Laser (DML) producing strong pulse confinement in each timeslot. The measured extinction ratio is 83.2~dB obtained by comparing the energies measured in the non-data to the data timeslots. The signal goes through Variable Optical Attenuators (VOAs) followed by a Polarization Controller (PC) before entering the SNSPD operating at 2.7K with a system detection efficiency of 85\% and 20 dark counts per second. The photon-counting events are assigned timestamps followed by clock and data recovery using offline signal processing. The PPM Frame Erasure Ratio (FER), PPM frame errors, and BER are extracted and achievable PIE are calculated. The overall system jitter was measured to be 45~ps at full width half maximum. Both frames without a photon and frames with two photons or more are considered erasure frames for the PIE calculation. 

\section{Clock recovery and synchronization}
Two common methods to perform clock recovery in PPM communication are either by inserting empty guard times or pilot PPM frames containing a pulse at a prescribed location. In both cases, clock recovery relies on statistical accumulation with the maximum PPM length and were previously demonstrated only for relatively short PPM frames with $M = 7$\cite{rogalin2016maximum,rielander2023esa} or smaller\cite{mendenhall2007design, hopman2006end}.

In this paper, clock timing is recovered by first performing clock pulse extraction, followed by clock ticks outliers removal and clock timing interpolation that closely tracks the clock random walk. No time averaging is required, reducing the clock sampling period and laser stability requirement.
\begin{figure}[ht]
   \centering
    \includegraphics[width=1\linewidth]{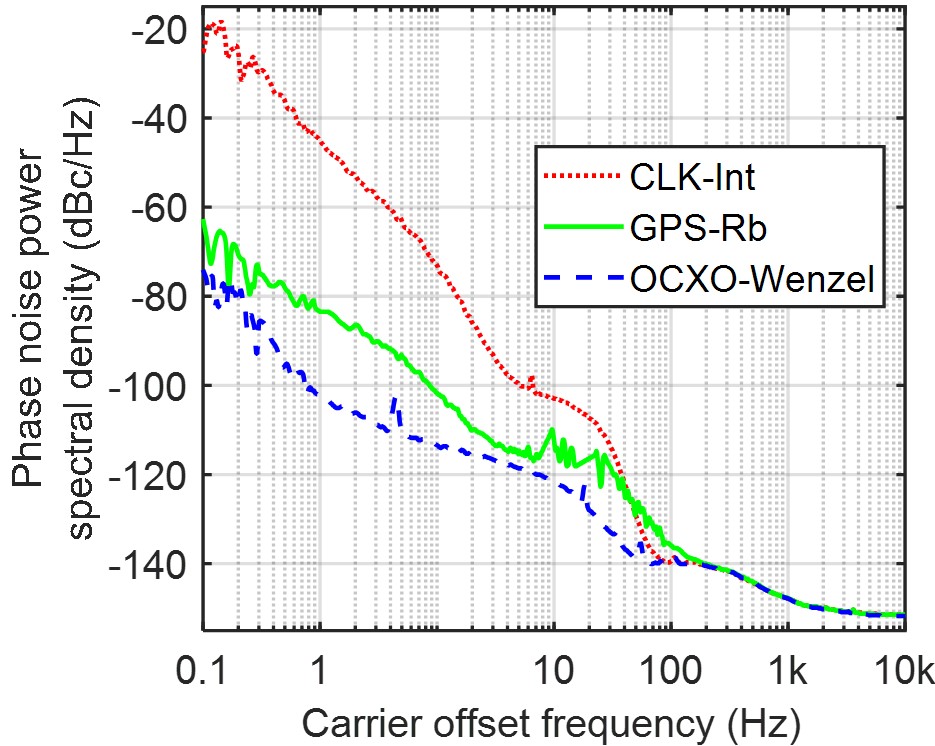}
    \caption{Measured phase noise power spectral density versus carrier offset frequency for three 10-MHz clocks at the output of the AWG; CLK-Int: from the 100-MHz AWG internal clock; GPS-Rb: from 10-MHz GPS-disciplined Rubidium clock; OCXO-Wenzel: from 10-MHz Wenzel model 501-27521-12.}
    \label{fig:figure2}
\end{figure}

The clock recovery was performed with a clock sampling period ranging from 50~ms to 2~seconds, corresponding to clock phase stability up to frequencies as low as 20~Hz to 0.5~Hz, respectively. Figure~\ref{fig:figure2} shows the measured phase noise power spectral density of the internal free-running clock of the AWG, CLK-Int, used in this paper, as well as a standard GPS-disciplined Rubidium (Rb) clock used in cellular base stations and a free-running oven-controlled crystal (Xtal) Oscillator (OCXO). The spectra of CLK-Int shows 8 to 35~dB worse phase noise in the 20 to 0.5 Hz range, evidence that moderate-quality free-running clocks without GPS disciplining are usable with this clock recovery scheme. 
\begin{figure}[ht]
   \centering
    \includegraphics[width=1\linewidth]{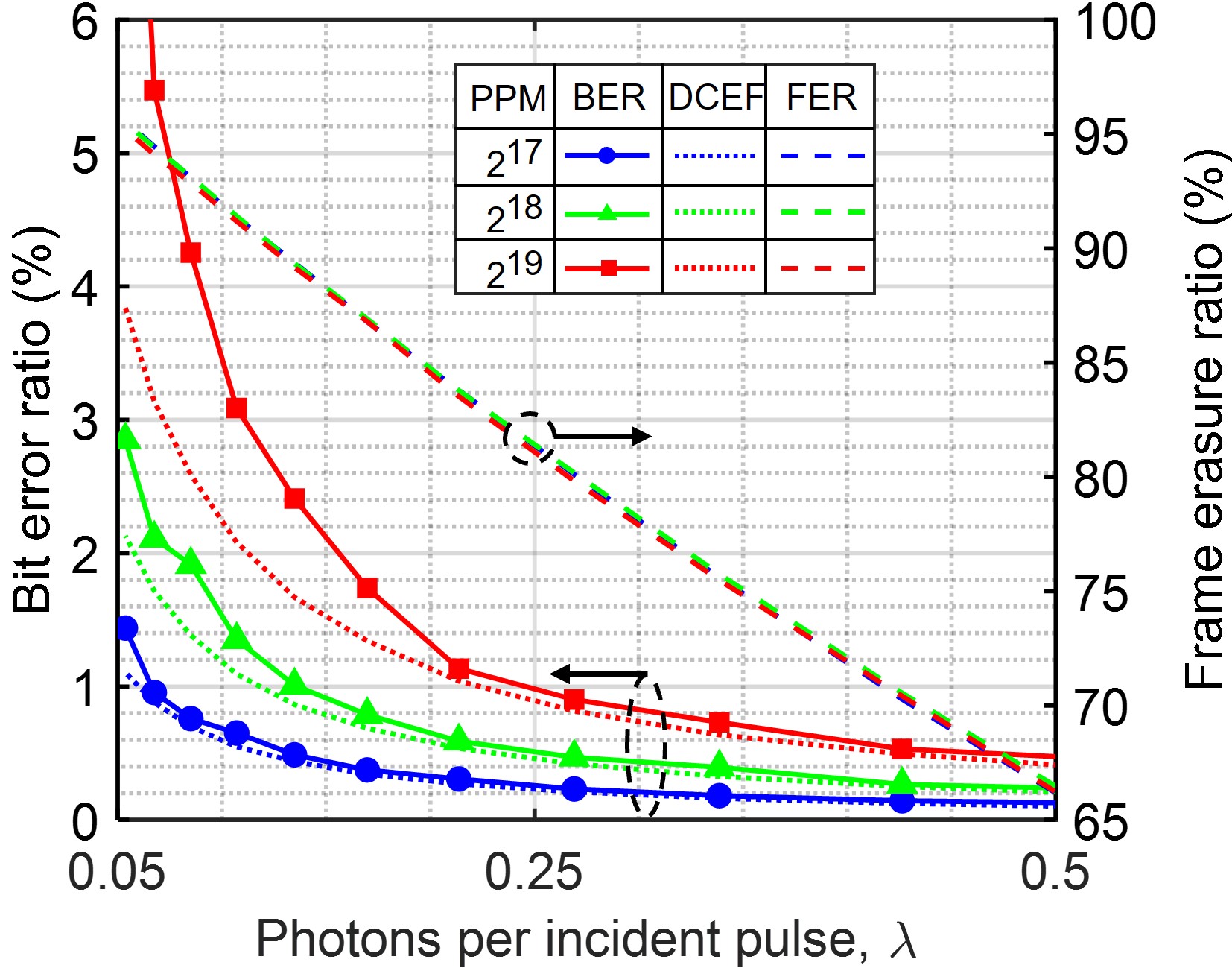}
    \caption{Measured bit error ratio (BER) and frame erasure ratio (FER) versus the number of photons per incident pulse, $\lambda$, for $2^{17}$,$2^{18}$ and $2^{19}$-PPM having a data-clock ratio $N = 10$. Also displayed is the BER contribution due to the Dark-Count on Erasured Frame (DCEF) as a guide.}
    \label{fig:figure3}
\end{figure}

\section{Measured photon information efficiency}
The measured BER for the non-erasured frames and the FER versus optical power are shown in Fig.~\ref{fig:figure3} for a data-to-clock ratio of $N=10$. The optical power is in the number of photons per incident pulse, $\lambda$, including both data and clock pulses. The PPM frame lengths are $2^{17}$, $2^{18}$ and $2^{19}$. The dotted lines are BERs originating solely from the Dark-Counts on Erased Frames (DCEFs), demonstrating that it is the dominant source of errors for $\lambda \ge 0.21$ photons per incident pulse. For $\lambda < 0.21$, the increased BER originates from additional errors introduced by the random walk of the clock that starts to appear. The FERs are represented by dashed lines and closely follow the Poissonian rate of frame erasures. 

\begin{figure}[ht]
   \centering
    \includegraphics[width=1\linewidth]{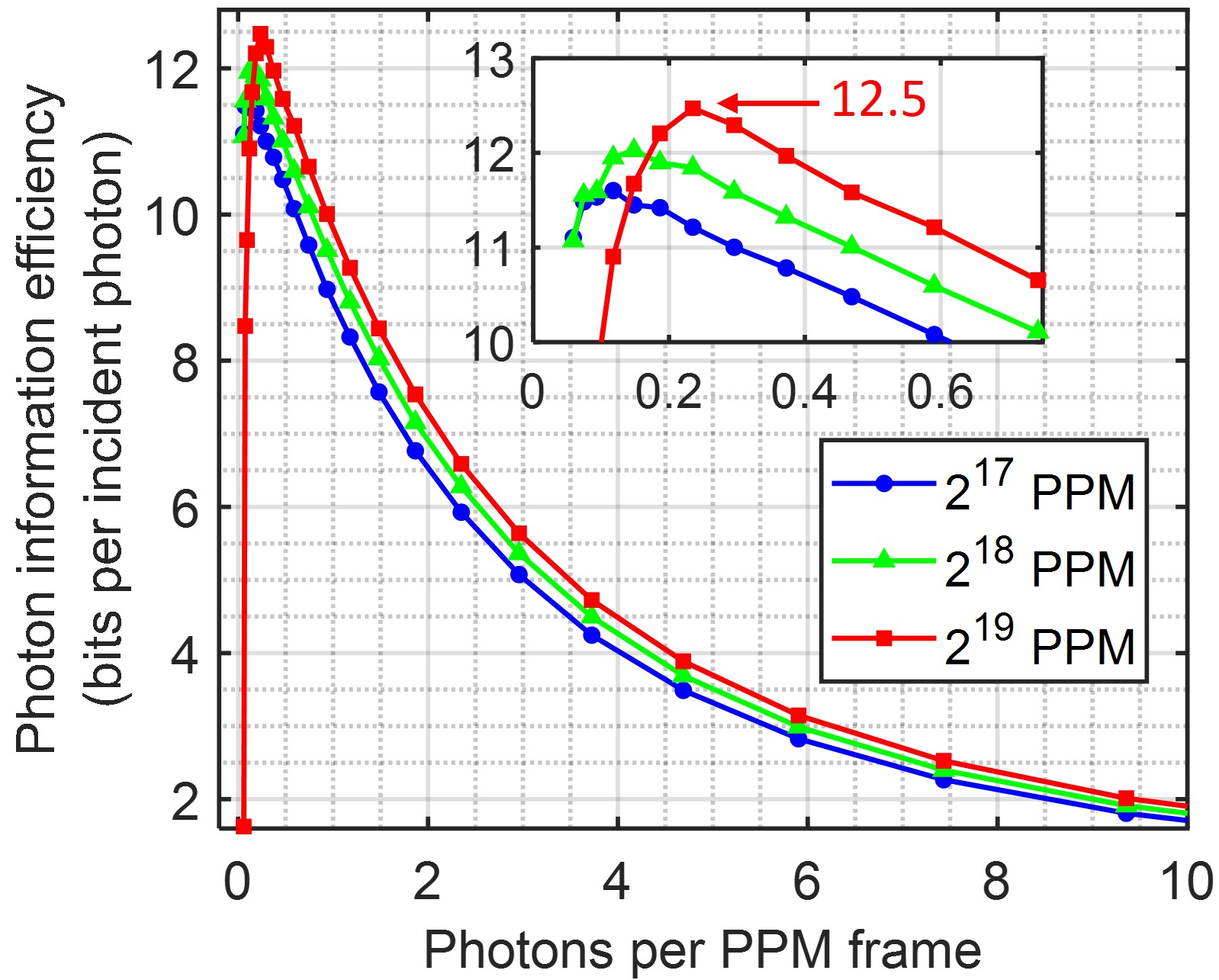}
    \caption{Photon information efficiency (in bits per incident photon, BIP) versus photons per PPM frame for $2^{17}$,$2^{18}$ and $2^{19}$-PPM with data-clock-ratio N = 10}
    \label{fig:figure4}
\end{figure}

For a channel experiencing both errors and erasures, a Reed-Solomon (RS) Forward-Error Correction (FEC) code is highly suitable. Figure~\ref{fig:figure4} shows the PIEs calculated from the measured frame error rates and FERs using an RS FEC code similar to the code implemented in~\cite{farr201313}. The photon per PPM frame is given by $(1+1/N)\,\lambda=1.1\,\lambda$. A measured achievable PIE of 12.5 BIP is obtained for $2^{19}$-PPM at a data rate of 13.94 kbits/s with photons per PPM frame of 0.235 (photons per pulse $\lambda=$0.213) with 0.154 RS FEC code rate at $10^{-6}$ decoded block error rate, an output block error rate typical for deep-space communication. The maximum PIE uncertainty is estimated to be $\pm$0.42 BIP, based on signal power measurement uncertainty. This measured PIE is 9.4 dB lower power than the $1/\ln(2) = 1.44$~BIP theoretical PIE limit of conventional dual-quadrature coherent detection\cite{gordon1962quantum} and 6.4~dB from the ultimate $2/\ln(2) = 2.88$~BIP single-quadrature coherent detection\cite{banaszek2020quantum}. This experiment demonstrates the feasibility of communication at high PIE ($>10$ BIP) with large PPM order ($>2^{10}$-PPM) and low photon flux ($< 0.1$~photon per PPM frame) with independent free-running clocks.
\section{Conclusions}


We reported transmission at a photon information efficiency of 12.5 bits per incident photon. To the best of our knowledge, this is the most power-efficient optical transmission demonstration to date with independent free-running clocks. 



\pagebreak
\printbibliography
\end{document}